\documentclass[11pt,a4paper]{article}
\usepackage{mathrsfs}
\usepackage{bm}
\usepackage{amsmath,amssymb}
\usepackage[dvipdfmx]{graphicx}
\usepackage[colorlinks]{hyperref}%arXiv
\usepackage[top=25truemm,bottom=25truemm,left=20truemm,right=20truemm]{geometry}
\title{Bending of Light and Inhomogeneous Picard-Fuchs Equation}
\author{Tadashi Sasaki${}^1$\footnote{t-sasaki@high.hokudai.ac.jp} and Hisao Suzuki${}^2$\footnote{hsuzuki@particle.sci.hokudai.ac.jp} \protect\\
	${}^1$Institute for the Advancement of Higher Education, \protect\\ Hokkaido University, Sappro 060-0817, Japan \protect\\
	${}^2$Department of Physics, Hokkaido University, Sapporo 060-0810, Japan}
\date{}
\begin{document}

\maketitle

\abstract{
 Bending of light rays by gravitational sources is one of the first evidences of the general relativity. 
 When the gravitational souce is a stationary massive object such as a black hole, the bending angle has an integral representation, from which
 various series expansions in terms of the parameters of orbit and the background spacetime has been derived.
 However, it is not clear that it has any analytic expansion.
 In this paper, we show that such an analytic expansion can be obtained for the case of a Schwarzschild black hole  by solving an inhomogeneous Picard-Fuchs equation, 
 which has been applied to compute effective superpotentials on D-branes in the Calabi-Yau manifolds.
 From the analytic expression of the bending angle, both weak and strong deflection expansions are explicitly obtained.
 We show that the result can be obtained by the direct integration approach.
 We also discuss how the charge of the gravitational source affects the bending angle and show that a similar analytic expression can be obtained
 for the extremal Reissner-Nordstr{\"{o}}m spacetime. }

%\begin{titlepage}
%\setcounter{page}{0}
%\begin{flushright}
%EPHOU 08-???\\
%February 2008\\
%\end{flushright}

\section{Introduction} 
% EHTとの関連を書く。
% optical appearance of a black hole \cite{Luminet,EHT}
  The theory of general relativity was proposed by Einstein in 1915. 
  One of the important predictions is bending of light ray in the presence of gravitational fields. 
  In particular, deflection by astrophysical gravitational sources such as stars, black holes, or galaxies has been studied both theoretically and observationally.

  \begin{equation}
	\alpha = \frac{4M}{b},
  \end{equation}
  where $M$ is the mass of the lensing object and $b$ is the impact parameter of the light trajectory.
  In the weak deflection limit, $b\gg M$ is assumed, which is satisfied for the lensing by a star such as the sun.
  Note that although deflection of light can be derived even in Newtonian gravity, the bending angle in general relativity is almost twice as much as that in Newton's theory. 
  This prediction was confirmed observationally in 1919 during the total solar eclipse\cite{Eddington}.

  When the lensing objects are very massive, it can give rise to a deviation from the above expression.
  The extension to include the higher order corrections in terms of the power series expansion with $M/b$ has been studied\cite{Keeton,Iyer}.
  Also, generalizations to the case with nonvanishing spin and electric charge are found for example in ref.\cite{ChakrabortySen}.

  On the other hand, in the strong deflection limit, i.e. the impact parameter and the Schwarzschild radius are comparable, 
  the light ray can wind around the object arbitrary times producing an infinite number of images, called relativistic images.
  This behavior is related to the circular orbit of photons, whose coordinate radius is $r=3M$ for the Schwarzschild spacetime. 
  Analytically, the existence of the relativistic images can be understood by observing the logarithmic divergence of $\alpha$ when the impact parameter $b$ approaches the critical value\cite{Darwin}.
  The strong deflection expansion of $\alpha$ beyond the leading divergence for the Schwarzschild case was performed in ref.\cite{Iyer}. 
  For the cases with spin and/or electric charge, only numerical calculations using expressions with various elliptic integrals can be found in the strong field limit\cite{Eiroa,Hsiao}.
  
  The strong deflection limit of the bending angle has been of interest also because it is relevant to the optical appearance, or the  shadow of a black hole\cite{Luminet},
  which has been recently observed by the Event Horizon Telescope\cite{EHT}.

  In deriving the results mentioned above, the starting point is usually expressions of the deflection angle $\alpha$ (more precisely, $\Theta:=\alpha+\pi$) in terms of the standard elliptic integrals,
  and the parameters of these integrals such as the modulus are complicated functions of the parameters of the background spacetime and the trajectory.
  Therefore we cannot read off from these expressions arbitrarily higer order terms in both weak and strong deflection limits 
  although the first few terms have been obtained.

  One of the aims of this paper is to obtain full order expansion in terms of $M/b$ of the deflection angle for the Schwarzschild case.
  Our strategy is to consider a differential equation satisfied by $\Theta$ as a function of $M/b$, 
  which can be seen as an inhomogeneous Picard-Fuchs equation. 
  Picard-Fuchs equations are differential equations with respect to the moduli of algebraic manifolds satisfied by the periods integrals.
  The notion of Picard-Fuchs equations has been used in physics.
  For example, it was applied to study the dependence of some Feynman integrals on the external variables\cite{PFeqFeynmanInt}.
  Also, it has been applied to analyze the dependence of D-brane superpotentials on the complex moduli in various Calabi-Yau manifolds 
  in the context of mirror symmetry\cite{Walcher1,KrefWalcher,KnappScheidegger,Fujietal,ShimizuSuzuki},
  and in these cases the differential equations arise with inhomogeneous terms.

  The differential equation for $\Theta$, which is derived in Sec.\ref{PFSchwarzschild} after introducing our notation in Sec.\ref{Schwarzschild}, 
  turns out to be a hypergeometric one with an inhomogeneous term.
  The explicit solution is given in terms of hypergeometric functions, from which the coefficients of arbitrarily higher order terms in the weak deflection limit can be easily read off. 
  By using analytic continuation formulas for the hypergeometric functions and the inhomogeneous Picard-Fuchs equation, 
  the strong deflection expansion is also derived.
  The result is completely consistent with the previous results in refs.\cite{Keeton,Iyer}.
  In Sec.\ref{directintegration}, the same result is rederived directly performing the defining integral of $\Theta$ with the help of analytic continuation to confirm our result.
  Furthermore, we consider a generalization of the method used in Sec.\ref{PFSchwarzschild} to the Reissner-Nordstr{\"{o}}m black hole, 
  and show that for the extremally-charged case $\Theta$ is given in a similar expression to the uncharged case in Sec.\ref{PFRN}.
  Finally, we conclude this paper with some discussions.

  %%%%%%%%%%%%%%%%%%%%%%%%%%%%%%%%%%%%
\section{Bending of light in the Schwarzschild geometry\label{Schwarzschild}}

\begin{figure}[htbp]
\begin{center}
\includegraphics[clip,height=4.0cm]{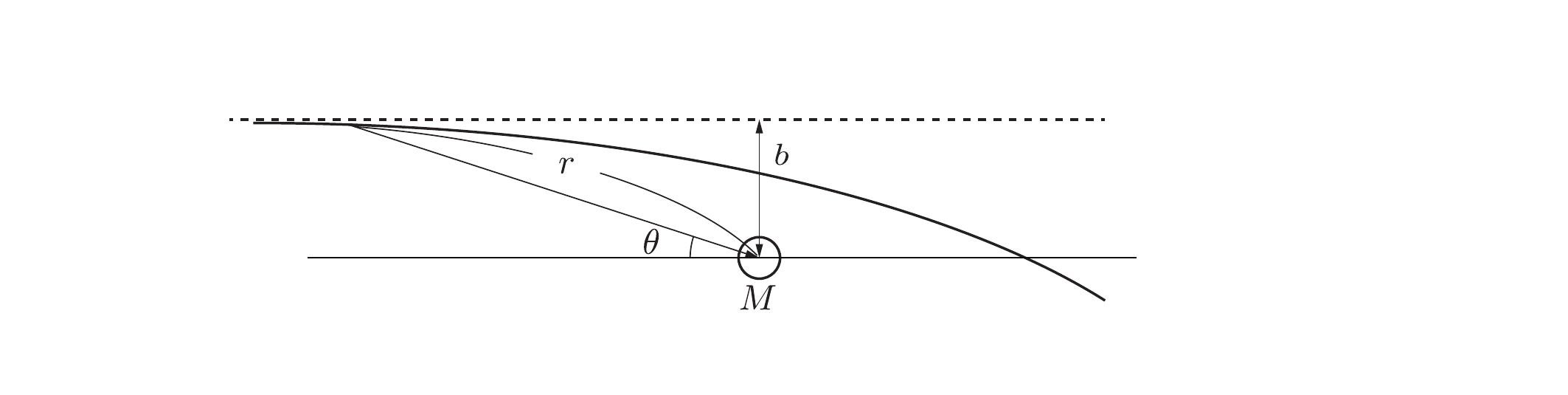}
\caption{Trajectory of light ray.}
\label{fig:orbit}
\end{center}
\end{figure}

The Action for a massless particle is given by
\begin{equation}
	S=\frac{N}{2}\int d\tau g_{\mu\nu}(x(\tau))\frac{dx^\mu}{d\tau}\frac{dx^\nu}{d\tau},\label{eq:particleaction}
\end{equation}
where $N$ is a Lagrangian multiplier field (we will set $N=1$ later).
We consider the Schwarzschild geometry
\begin{equation}
	ds^2=-\left(1-\frac{2M}{r}\right)dt^2+\left(1-\frac{2M}{r}\right)^{-1}dr^2+r^2d\theta^2+r^2\sin\theta^2d\phi^2
\end{equation}
and (\ref{eq:particleaction})is written as
\begin{equation}
	S=\frac{N}{2}\int d\tau \left[-\left(1-\frac{2M}{r}\right)\left(\frac{dt}{d\tau}\right)^2+\left(1-\frac{2M}{r}\right)^{-1}\left(\frac{dr}{d\tau}\right)^2+r^2\left(\frac{d\theta}{d\tau}\right)^2+r^2\sin\theta^2\left(\frac{d\phi}{d\tau}\right)^2\right].
\end{equation}
The variation with respect to $N$ and setting $N=1$ leads to the null condition,
\begin{equation}
	-\left(1-\frac{2M}{r}\right)\left(\frac{dt}{d\tau}\right)^2+\left(1-\frac{2M}{r}\right)^{-1}\left(\frac{dr}{d\tau}\right)^2+r^2\left(\frac{d\theta}{d\tau}\right)^2+r^2\sin\theta^2\left(\frac{d\phi}{d\tau}\right)^2=0. \label{eq:Nequation}
\end{equation}
Variation with respect to $t,\theta,\phi$ leads to
\begin{equation}
	(1-\frac{2M}{r})\frac{dt}{d\tau}=\varepsilon,\qquad r^2\frac{d\theta}{d\tau}=L,\qquad r^2\sin^2\theta \frac{d\phi}{d\tau}=m,\label{eq:elm}
\end{equation}
where $\varepsilon,L,m$ are constants.
By setting the polar coordinates, we set $\phi=0\ (m=0)$ and inserting (\ref{eq:elm}) into (\ref{eq:Nequation}), we find
\begin{equation}
	\frac{1}{L^2}\left(\frac{dr}{d\tau}\right)^2+\left(1-\frac{2M}{r}\right)\frac{1}{r^2}=\frac{1}{b^2},
\end{equation}
where $b=L/\varepsilon$.
We will write the trajectory as a function of $\theta$, we find
\begin{equation}
	\frac{1}{r^4}\left(\frac{dr}{d\theta}\right)^2+\left(1-\frac{2M}{r}\right)\frac{1}{r^2}=\frac{1}{b^2}.
\end{equation}
By setting the variable $x$ as
\begin{equation}
	x=\frac{1}{r}
\end{equation}
we find
\begin{equation}
	\left(\frac{dx}{d\theta}\right)^2=-(1-2Mx)x^2+\frac{1}{b^2}.\label{eq:firstorder}
\end{equation}
Note that at far distant region ($\Leftrightarrow x\ll1$) where the gravitational field of the source can be negligible, 
the solution of the equation can be approximated by $r\sin\theta=b$. 
Therefore $b$ is the impact parameter of the photon trajectory as can be seen in Fig.\ref{fig:orbit}.
Equation (\ref{eq:firstorder}) is the first order equation and we can easily obtain the integral form of the deflection angle $\Theta$ as
\begin{equation}
	\Theta=2\int_0^{x_0}\frac{dx}{\sqrt{1/b^2-x^2+2Mx^3}}, \label{angle} %deflection angle \Theta_{\rm{def}}=\Theta-\pi
\end{equation}
where $x_0$ is the turning point of the orbit and given by one of the roots of the equation $1/b^2-x^2+2Mx^3=0$, which reduces to $x=1/b$ when $M=0$.

We can perturbatively obtain the series of expansion 
\begin{equation}
	\Theta=\pi+\frac{4M}{b}+\frac{15\pi}{4}\left(\frac{M}{b}\right)^2+\frac{128}{3}\left(\frac{M}{b}\right)^3+\frac{3465\pi}{64}\left(\frac{M}{b}\right)^4+O((M/b)^5),\label{Keetonresult}
\end{equation}
which was obtained in ref.\cite{Keeton}\footnote{In ref.\cite{Keeton}, the explicit coefficients are given up to the order of $(M/b)^6$.}.
Our aim of this paper is to obtain the series of the expansion for all order.
We change the variable of integration to $t=bx$ and set the constants $\alpha, \beta$, and $\gamma$ so that 
\begin{equation}
	1-t^2+\frac{2M}{b}t^3=(1-\alpha t)(1-\beta t)(1-\gamma t).
\end{equation}
The photon can reach the observer, which is assumed to be at the infinity, when the impact parameter is sufficiently large, i.e. $2M/b\leq(4/27)^2$.
In this case, the roots of the above equation are all real and we can take the constants so that $\gamma<0<\beta<\alpha$.
Then we get
\begin{equation}
	\Theta=2\int_0^{1/\alpha} \frac{dt}{(1-\alpha t)^{\frac{1}{2}}(1-\beta t)^{\frac{1}{2}}(1-\gamma t)^{\frac{1}{2}}}
\end{equation}
Changing the variable to $s=\alpha t$, we have
\begin{equation}
\Theta=\frac{2}{\alpha}\int_0^1 ds\frac{1}{(1-s)^{\frac{1}{2}}(1-\frac{\beta}{\alpha}s)^{\frac{1}{2}}(1-\frac{\gamma}{\alpha} s)^{\frac{1}{2}}}.
\end{equation}
Taking the power series and making the integration, we get
\begin{align}%ここの数式は要チェック-->チェック済み
	\Theta=&\frac{2}{\alpha}\sum_{m,n=0}^\infty\frac{(1/2)_m(1/2)_n}{m!n!}\frac{\Gamma(m+n+1)\Gamma(1/2)}{\Gamma(m+n+\frac{3}{2})}\left(\frac{\beta}{\alpha}\right)^m\left(\frac{\gamma}{\alpha}\right)^n\notag\\
	=&\frac{4}{\alpha}F_1\left(1,\frac{1}{2},\frac{1}{2};\frac{3}{2};\frac{\beta}{\alpha},\frac{\gamma}{\alpha}\right), \label{Appelleq}
\end{align}
where $F_1$ is the Appell function\cite{HTF} and the pochhammer symbol is defined as $(x)_n:=\Gamma(x+n)/\Gamma(x)$. 
However, $\alpha,\beta$, and $\gamma$ are complicated functions of $M/b$ so this expression is not adequate to obtain the series expansion in terms of $M/b$.
In the next section, we try a different approach.

%%%%%%%%%%%%%%%%%%%%%%%%%%%%%%
\section{Inhomogeneous Picard-Fuchs equation\label{PFSchwarzschild}}

The integral (\ref{angle}) can be written as an incomplete elliptic integral.  %t integralで明記しておくべきか
It is related to the algebraic curve
\begin{equation}
	y^2=1-t^2+\frac{2M}{b}t^3,
\end{equation}
which is the defining equation of an algebraic torus in the complex $t-y$ plane.
Holomorphic one-form on the torus can be defined as
\begin{align}
	\omega=&\oint \frac{dy \wedge dt}{2\pi i}\frac{1}{y^2-(1-t^2+\frac{2M}{b}t^3)}\notag\\ %係数確認。というか1行目不要？
	=&\frac{dt}{\sqrt{1-t^2+\frac{2M}{b}t^3}}
\end{align}
It is known that the holomorphic one-form $\omega$ satisfies the Picard-Fuchs equation with respect to the moduli $2M/b$.
Let the Picard-Fuchs operator be $D(\partial_z)$ (the definition of $z$ is given soon).
Then the holomorphic one-form $\omega$ satisfies 
\begin{equation}
	D(\partial_z)\omega=-d\beta,
\end{equation}
where $\beta$ is a zero form. 
In fact, the Picard-Fuchs operator and the zero-form $\beta$ can be obtained by this requirement.
Taking the cyclic integral of this identity, the Picard-Fuchs equation can be obtained,
\begin{equation}
	D(\partial_z)\oint d\omega=0.
\end{equation}
The integral (\ref{angle}) can be considered as the integral with the boundary $t=0$. 
Therefore acting $D$ will leads to the inhomogeneoust Picard-Fuchs equation,
\begin{equation}
	D(\partial_z)\int d\omega=\beta(t=0).
\end{equation}
Now let us define the moduli parameter $z$
\begin{equation}
	\frac{M}{b}=\frac{1}{3^\frac{3}{2}}z^{\frac{1}{2}}
\end{equation}
and the diffential operator $\theta=z\frac{d}{dz}$. 

We find that the deflection angle satisfies the following Picard-Fuchs equation:
\begin{equation}
	P\Theta=\frac{1}{3^\frac{3}{2}}z^{\frac{1}{2}}, \label{ihPFeq}
\end{equation}
where the operator $P$ is given by
\begin{equation}
	P=\theta^2-z\left(\theta+\frac{1}{6}\right)\left(\theta+\frac{5}{6}\right).
\end{equation}
It is easy to get the following special solution of this equation:
\begin{equation}
	\frac{4}{3^\frac{3}{2}}z^\frac{1}{2}{}_3F_2\left[\genfrac{}{}{0pt}{0}{2/3,1,4/3}{3/2,3/2};z\right].
\end{equation}
Two independent solutions of the homogeneous equation are hypergeometric functions ${}_2F_1[1/6,5/6;1;z]$ 
and ${}_2F^*_1[1/6,5/6;1;z]:={}_2F_1[1/6,5/6;1;1-z]$, 
the latter of which contains the logarimic singularity around $z=0$.
The general solution is then given by
\begin{equation}
	\Theta=c_1\cdot {}_2F_1\left[\frac{1}{6},\frac{5}{6};1;z\right]+c_2\cdot {}_2F^*_1\left[\frac{1}{6},\frac{5}{6};1;z\right]+\frac{4}{3^\frac{3}{2}}z^\frac{1}{2}{}_3F_2\left[\genfrac{}{}{0pt}{0}{2/3,1,4/3}{3/2,3/2};z\right],
\end{equation}
where $c_1$ and $c_2$ are integration constants.
Since the orbit becomes a straight line in the limit $z\to0$, i.e. $\Theta(z=0)=\pi$,
these constants are fixed uniquely,
\begin{equation}
	\Theta=\pi {}_2F_1\left[\frac{1}{6},\frac{5}{6};1;z\right]+\frac{4}{3^\frac{3}{2}}z^\frac{1}{2}{}_3F_2\left[\genfrac{}{}{0pt}{0}{2/3,1,4/3}{3/2,3/2};z\right],
\end{equation}
which can be expressed by the original variables as
\begin{align}
	\Theta&=\pi {}_2F_1\left[\frac{1}{6},\frac{5}{6};1;\frac{27M^2}{b^2}\right]+\frac{4M}{b}{}_3F_2\left[\genfrac{}{}{0pt}{0}{2/3,1,4/3}{3/2,3/2};\frac{27M^2}{b^2}\right], \label{Schwarzschildsolution}\\
	&=\pi\sum_{n=0}^\infty\frac{(1/6)_n(5/6)_n}{(n!)^2}\left(\frac{27M^2}{b^2}\right)^n+\frac{4M}{b}\sum_{n=0}^\infty\frac{(2/3)_n(4/3)_n}{((3/2)_n)^2}\left(\frac{27M^2}{b^2}\right)^n.
\end{align}
It is easy to check that the first few series expansion agrees with (\ref{Keetonresult}).%ref必要-->入れた

To understand the strong deflection limit of this solution, we need analytic continuation formulae for hypergeometric functions.
For ${}_2F_1$, such an identity is a classical result\cite{HTF},
\begin{equation}
	{}_2F_1\left[a,b;a+b;z\right]=\frac{\Gamma(a+b)}{\Gamma(a)\Gamma(b)}\sum_{n=0}^\infty
	\frac{(a)_n(b)_n}{(n!)^2}\left[k_n-\log(1-z)\right](1-z)^n, \label{2f1formula}
\end{equation}
where $k_n=2\psi(n+1)-\psi(n+a)-\psi(n+b)$ with $\psi(z)=\Gamma'(z)/\Gamma(z)$ being the digamma function.
For ${}_3F_2$, we can use the following identity given in ref.\cite{Wolfgang}\footnote{$-\psi(n+a_1)$ is missing in Eq.(5.1) in this refference article.}
\begin{align}
	&\frac{\Gamma(a_1)\Gamma(a_2)\Gamma(a_3)}{\Gamma(b_1)\Gamma(b_2)}{}_3F_2\left[\genfrac{}{}{0pt}{0}{a_1,a_2,a_3}{b_1,b_2};z\right] \notag\\
	=&\sum_{n=0}^\infty\frac{(a_1)_n(a_2)_n}{(n!)^2}\left\{\sum_{k=0}^n\frac{(-n)_k}{(a_1)_k(a_2)_k}A_k^{(2)}\left(\psi(n-k+1)+\psi(n+1)\right.\right.\notag\\
	&\hspace{6em}\left.-\psi(n+a_1)-\psi(n+a_2)-\log(1-z)\right)\notag\\
	&\hspace{4em}\left.
	+(-1)^nn!\sum_{k=n+1}^\infty\frac{(k-n-1)!}{(a_1)_k(a_2)_k}A_k^{(2)}\right\}(1-z)^n, \label{3f2formula}
\end{align}
where $A_k^{(2)}=(b_2-a_3)_k(b_1-a_3)_k/k!$.
Note that this formula is valid only when ${}_3F_2$ is the so-called zero-balanced series, meaning that the parameters satisfy $a_1+a_2+a_3=b_1+b_2$.

Although these formulas are enough to understand the strong deflection limit, 
the inhomogeneous Picard-Fuchs equation (\ref{ihPFeq}) helps us simplify the expansion about $w:=1-z$ as we show now.
Observe that $\Theta$ takes the following form:
\begin{equation}
	\Theta(w)=p(w)+q(w)\log w,
\end{equation}
where $p(w)$ and $q(w)$ have power series expansions around $w=0$.
Since the inhomogeneous term in Eq.(\ref{ihPFeq}) is free of logarithmic singularity $\log w$, $q(w)$ must be a homogeneous solution which is regular at $w=0$.
As a result, $q(w)$ is proportional to ${}_2F_1[1/6,5/6;1;w]$.
By inserting Eqs.(\ref{2f1formula}) and (\ref{3f2formula}) into the expression (\ref{Schwarzschildsolution}) and picking up $(n=0)$-term,
we can identify the proportionality constant to conclude
\begin{equation}
	q(w)=-{}_2F_1\left[\frac{1}{6},\frac{5}{6};1;w\right].
\end{equation}

$p(w)$ is determined by solving the inhomogeneous hypergeometric equation assuming a power series expansion around $w=0$,
but it is more convenient to define $\tilde{p}(w)$ as follows:
\begin{equation}
	\Theta(w)=\tilde{p}(w)+2\pi {}_2F_1\left[\frac{1}{6},\frac{5}{6};1;1-w\right],
\end{equation}
which is equivalent to $p(w)=\tilde{p}(w)+r(w)$, where 
\begin{equation}
	r(w)=\sum_{n=0}^\infty\frac{\left(\frac{1}{6}\right)_n\left(\frac{5}{6}\right)_n}{(n!)^2}\left(2\psi(n+1)-\psi(n+1/6)-\psi(n+5/6)\right)w^n.
\end{equation}
Then, $\tilde{p}(w)$ is the solution of $P(p)=\sqrt{1-w}/3\sqrt{3}$ with a power series expansion $\tilde{p}(w)=\sum\tilde{p}_nw^n$,
of which coefficients obey the following reccurrence relation:
\begin{equation}
	n^2\tilde{p}_n=\left(n-\frac{1}{6}\right)\left(n-\frac{5}{6}\right)\tilde{p}_{n-1}+\frac{1}{3\sqrt{3}}\frac{\left(\frac{1}{2}\right)_{n-1}}{(n-1)!}, \ \ n\geq1.
\end{equation}
The general solution for $n\geq1$ is given by
\begin{equation}
	\tilde{p}_n=\frac{\left(\frac{1}{6}\right)_n\left(\frac{5}{6}\right)_n}{(n!)^2}\tilde{p}_0+\frac{\left(\frac{7}{6}\right)_{n-1}
		\left(\frac{11}{6}\right)_{n-1}}{(n!)^2}\sum_{j=0}^{n-1}\frac{j!\left(\frac{1}{2}\right)_j}{\left(\frac{7}{6}\right)_j
		\left(\frac{11}{6}\right)_j},
\end{equation}
where $\tilde{p}_0$ is an arbitrary constant.
The first term gives a contribution $\tilde{p}_0{}_2F_1[1/6,5/6;1;w]$.
Similarly, by using the identity $\psi(x+1)=\psi(x)+1/x$ we can subtract the hypergeometric function from $r(w)$ as
\begin{align}
	r(w)&=(2\psi(1)-\psi(1/6)-\psi(5/6)){}_2F_1\left[\frac{1}{6},\frac{5}{6};1;w\right]\notag\\
	&\hspace{2em}
		+\sum_{n=1}^\infty\frac{\left(\frac{1}{6}\right)_n\left(\frac{5}{6}\right)_n}{(n!)^2}
		\sum_{j=1}^n\left(\frac{2}{j}-\frac{1}{j-1/6}-\frac{1}{j-5/6}\right)w^n.
\end{align}
The digamma function at rational numbers can be expressed in terms of elementary functions\cite{HTF} so that the coefficient of the hypergeometric 
function is given by
\begin{equation}
	2\psi(1)-\psi(1/6)-\psi(5/6)=\log 432.
\end{equation}
Combining above results, we obtain
\begin{align}
	\Theta&=\left(\tilde{p}_0+\log\frac{432}{w}\right){}_2F_1\left[\frac{1}{6},\frac{5}{6};1,w\right] \notag\\
	&\hspace{2em}
		+\sum_{n=1}^\infty\frac{\left(\frac{1}{6}\right)_n\left(\frac{5}{6}\right)_n}{(n!)^2}
		\sum_{j=1}^n\left(\frac{2}{j}-\frac{1}{j-1/6}-\frac{1}{j-5/6}-\frac{2}{3\sqrt{3}}\frac{(j-1)!\left(-\frac{1}{2}\right)_j}
		{\left(\frac{1}{6}\right)_j\left(\frac{5}{6}\right)_j}\right)w^n. \label{wexpansion}
\end{align}
The remaining task is to determine $\tilde{p}_0$.
From the analytic continuation formula for ${}_2F_1$ and ${}_3F_2$ given above, one of the expressions for $\tilde{p}_0$ is found to be
\begin{equation}
	%\tilde{p}_0+\log432=2\psi(1)-\frac{1}{2}(\psi(1/6)+\psi(5/6)+\psi(2/3)+\psi(4/3))+\frac{9}{64}{}_4F_3\left[
	%\genfrac{}{}{0pt}{0}{1,1,3/2,3/2}{5/3,2,7/3};1\right].
	\tilde{p}_0=-\log4-\frac{3}{2}+\frac{9}{64}{}_4F_3\left[
	\genfrac{}{}{0pt}{0}{1,1,3/2,3/2}{5/3,2,7/3};1\right].
\end{equation}
Another expression can be derived by using the method of variation of constant to obtain an integral expression for $\Theta$ in terms of 
the homogeneous solutions.
After some calculations we found
\begin{equation}
	\tilde{p}_0=-\frac{2\pi}{3\sqrt{3}}\int_0^1\frac{dz}{\sqrt{z}}{}_2F_1\left[\frac{1}{6},\frac{5}{6};1;z\right]
		=-\frac{4\pi}{3\sqrt{3}}{}_3F_2\left[\genfrac{}{}{0pt}{0}{1/6,1/2,5/6}{1,3/2};1\right]. \label{p0-3f2}
\end{equation}
In fact, one can directly expand the original expression of $\Theta$ to obtain
\begin{equation}
	\tilde{p}_0=\log(7-4\sqrt{3}). \label{p0-log}
\end{equation}
The equality of Eqs.(\ref{p0-3f2}) and (\ref{p0-log}) can be proved by using the Watson's formula\cite{Watson}  with the help
of contiguous relations\cite{Rainville} for ${}_3F_2$.
In \cite{Iyer}, a strong deflection limit of $\hat{\alpha}:=\Theta-\pi$ was considered and a first few coefficient in $b'$-expansion 
was obtained, where $b'=1-\sqrt{1-w}$.
From our result Eq.(\ref{wexpansion}), the expansion given in ref.\cite{Iyer} is completely recovered.

%To understand the strong deflection limit of $\Theta$, we consider the expansion in terms of $b'=1-\sqrt{z}$, which is used in ref.\cite{Iyer}.
%For that purpose, one of the quadratic transformations for ${}_2F_1$\cite{HTF},
%\begin{equation}
%	{}_2F_1[a,b;a+b+1/2;4x(1-x)]={}_2F_1[2a,2b;a+b+1/2;x],
%\end{equation}
%with $x=b'/2, a=1/6, b=5/6$

\section{Direct evaluation of the integral via analytic continuation\label{directintegration}}

\begin{figure}[htbp]
\begin{center}
\includegraphics[clip,height=4.0cm]{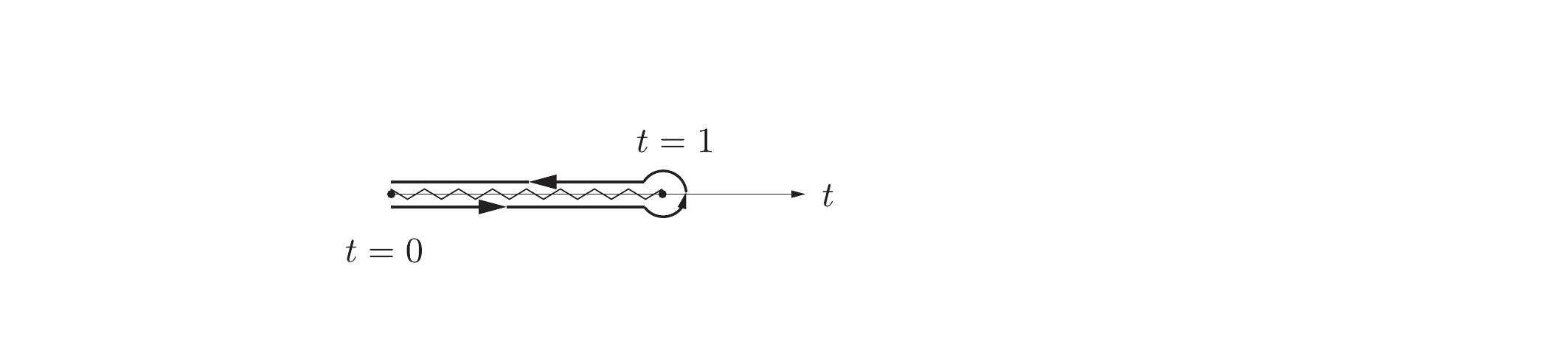}
\caption{The contour of $t$ which can be evaluated by 2 times the line integral from$ t=0$ to $t=1$.}
\label{fig:contour}
\end{center}
\end{figure}

In the previous section, we have obtained the deflection angle by using the inhomogeneous Picard-Fuchs equation.
The next natural question is whether we can obtain the result by direct integration.
In ref., the integral of the holomorphic forms for planes with boundaries has been performed via analytic continuation. %ref必要
We will apply the method.
Note that the integral can be written as
\begin{equation}
	\Theta=\int_C\frac{dt}{(1-t^2+\frac{2M}{b}t^3)^{\frac{1}{2}}},
\end{equation}
where $C$ is the line starting from 0 and encircling around the root and coming back to 0 (Fig.\ref{fig:contour}).
We will use the following representation
\begin{equation}
\frac{1}{(1-t^2+\frac{2M}{b}t^3)^{\frac{1}{2}}}=\int ds\Gamma(-s)(\frac{2M}{b})^s t^{3s}\frac{\Gamma(s+\frac{1}{2})}{\Gamma(\frac{1}{2})}(1-t^2)^{-s+\frac{1}{2}-1},
\end{equation}
where $s$ takes the pole of non-negative integers.
The original integral have a cut structure at the pole but now it has a cut structure at $t=1$. Therefore the integral can be evaluated by two times the line integral from 0 to 1.
Namely,
\begin{equation}
\Theta=2\int ds\Gamma(-s)(\frac{2M}{b})^s\frac{\Gamma(s+\frac{1}{2})}{\Gamma(\frac{1}{2})}\int_0^1 dt  t^{3s}(1-t^2)^{-s+\frac{1}{2}-1},
\end{equation}
We take line integral by using beta integral and find
\begin{equation}
\Theta=2\int ds\Gamma(-s)(\frac{2M}{b})^s\frac{\Gamma(\frac{3}{2}s+\frac{1}{2})}{\Gamma(\frac{1}{2})\Gamma(\frac{1}{2}s+1)\cos \pi s}.
\end{equation}
Taking the pose of $s$, we get
\begin{equation}
\Theta=\pi\sum_{n=0}^\infty \frac{(\frac{1}{2})_{\frac{3}{2}n}}{(1)_n(1)_{\frac{1}{2}n}}(\frac{2M}{b})^n
\end{equation}
This reslult coincide with (\ref{Schwarzschildsolution}).

\section{Photon Trajectories for extremal Reissner-Nordstr{\"{o}}m Geometry\label{PFRN}}

We have discussed the bending angle for Shwartzshild geometry. 
As we easily expected, the bending angle glows as the mass $M$ increases with the impact parameter $b$ kept fixed. 
The next question we consider is the effect of the charge for the bending angles. 
If the spherical object have electric charge, the spacetime metric is described by the Reissner-Nordstr{\"{o}}m solution. 
We can apply our direct integration method even for this geometry but it will lead an expansion with two variables.  
However, such an expression will not be so illuminating. 
Instead, we apply the method used in Sec.\ref{PFSchwarzschild} to the charged case.
%Here we will consider the photon trajectories in the extreme Reissner-Nordstrom space-time.

The Reissner-Nordstr{\"{o}}m metric is given by
\begin{equation}
	ds^2=-\left(1-\frac{2M}{r}+\frac{Q}{r^2}\right)dt^2+\left(1-\frac{2M}{r}+\frac{Q}{r^2}\right)^{-1}dr^2+r^2\theta^2+r^2\sin^2\theta d\phi^2,
\end{equation}
where $m$ is the mass of the star and $Q$ is the charge.
In this case, photon trajectories are described by 
\begin{equation}
	\frac{1}{r^4}\left(\frac{dr}{d\theta}\right)^2+\left(1-\frac{2M}{r}+\frac{Q^2}{r^2}\right)\frac{1}{r^2}=\frac{1}{b^2}.
\end{equation}
Introducing dimensionless parameters $u=2M/r$ and $q=Q/(2M)$, this equation can be reduced to the following elliptic form:
\begin{equation}
	\left(\frac{du}{d\theta}\right)^2=-q^2u^4+u^3-u^2+\frac{4M^2}{b^2},
\end{equation}
Before integrating this equation, we have to specify the parameter region of our interest.
In order that the photon can reach the observer at infinity without crossing the event horizon, the parameters must obey
\begin{equation}
	\frac{4M^2}{b^2}<f(u_-),\ \ f(u):=q^2u^4-u^3+u^2,
\end{equation}
where $u_\pm=(3\pm\sqrt{9-32q^2})/(8q^2)$ are the positions of extrema of $f(u)$.
In this case, all the roots $u_i (i=1,2,3,4)$ of $4M^2/b^2=f(u)$ are real so that we can assume $u_1<0<u_2<u_3<u_4$.
With this convention, integration of the differential equation gives the integral expression of the deflection angle $\Theta$,
\begin{equation}
	\Theta=2\int_0^{u_2}\frac{du}{\sqrt{-q^2u^4+u^3-u^2+4M^2/b^2}}=2\int_0^{u_2}\frac{du}{\sqrt{q^2(u-u_1)(u_2-u)(u_3-u)(u_4-u)}}.
\end{equation}

In addition to the parameter $M^2/b^2$, $\Theta$ depends on the squared background charge $q^2$.
As a result, the Picard-Fuchs equation now becomes a partial differential equation which contains a term proportional to $\partial_{q^2}\Theta$.
Interestingly, the coefficient of the additional term is proportional to $q^2(q^2-1/4)$, meaning the Picard-Fuchs equation reduces to 
ordinary differential equations when the background spacetime is the Schwarzschild one or the extremal Reissner-Nordstr{\"{o}}m one.
Hereafter we focus on the latter case, namely $q^2=1/4$.
In this case, the differential equation again turns out to be an inhomogeneous hypergeometric equation,
\begin{equation}
	\left[x(1-x)\frac{d^2}{dx^2}+(1-2x)\frac{d}{dx}-\frac{3}{16}\right]\Theta=\frac{1}{4\sqrt{x}},
\end{equation}
where the independent variable is differently normalized, $x=16M^2/b^2$.
The solution satisfying the boundary condition $\Theta(x=0)=\pi$ is uniquely identified as
\begin{equation}
	\Theta=\pi{}_2F_1\left[\frac{1}{4},\frac{3}{4};1;x\right]
	+\sqrt{x}{}_3F_2\left[\genfrac{}{}{0pt}{0}{3/4,1,5/4}{3/2,3/2};x\right],
\end{equation}
or in terms of the original variables,
\begin{equation}
	\Theta=\pi{}_2F_1\left[\frac{1}{4},\frac{3}{4};1;\frac{16M^2}{b^2}\right]+\frac{4M}{b}{}_3F_2\left[\genfrac{}{}{0pt}{0}{3/4,1,5/4}{3/2,3/2};\frac{16M^2}{b^2}\right]. \label{eRNsolution}
\end{equation}
Explicit coefficients up to 4th order terms are given by
\begin{equation}
	\Theta=\pi+\frac{4M}{b}+\frac{3\pi M^2}{b^2}+\frac{80M^3}{b^3}+\frac{105\pi M^4}{b^4}+O((M/b)^5),
\end{equation}
which is consistent with the previous result in ref.\cite{ChakrabortySen}.

As in the case of Schwarzschild spacetime, strong deflection limit of $\Theta$ can be derived.
We here show only the result,
\begin{align}
	\Theta&=\sqrt{2}{}_2F_1\left[\frac{1}{4},\frac{3}{4};1;y\right]\log\left(\frac{64(3-2\sqrt{2})}{y}\right) \notag\\
	&\hspace{2em}
	+\sum_{n=1}^\infty\frac{\left(\frac{1}{4}\right)_n\left(\frac{3}{4}\right)_n}{(n!)^2}\sum_{j=1}^n\left\{\sqrt{2}\left(\frac{2}{j}
	-\frac{1}{j-1/4}-\frac{1}{j-3/4}\right)+\frac{4}{3}\frac{(j-1)!\left(\frac{1}{2}\right)_{j-1}}{\left(\frac{5}{4}\right)_{j-1}
	\left(\frac{7}{4}\right)_{j-1}}\right\}y^n, \label{yexpansion}
\end{align}
where $y=1-x$.
In ref.\cite{Eiroa}, the leading logarithmic divergence of $\alpha = \Theta-\pi$ for generic value of electric charge was numerically studied.
There, they assumed the asymptotic form $\alpha=-A\log(B\epsilon)-\pi$, where $A$ and $B$ are constants depending on $q$ and $\epsilon$ represents 
deviation of the closest approach distance normalized by the Schwarzschild radius $x_0=r_0/2M$ from the photon sphere $x_{\rm ps}$, 
i.e. $x_0=x_{\rm ps}+\epsilon$.
In the strong deflection limit, $\epsilon$ and the variable $y$ are related as $y=2\epsilon^2$, which, with the help of Eq.(\ref{yexpansion}), gives
\begin{equation}
	\Theta=-2\sqrt{2}\log\frac{\sqrt{2}+1}{2^{5/2}}\epsilon+o(1),\ \ \epsilon\to0.
\end{equation}
This is completely consistent with the numerical result in ref.\cite{Eiroa} (see TABLE I).

\section{Summary and Discussions}

So far we have derived exact and explicit expressions of the bending angles of photon trajectories in Schwarzschild and extremal Reissner-Nordstr{\"{o}}m 
spacetimes in terms of the impact parameter by means of solving inhomogeneous Picard-Fuchs equations.
Our results are generalizations of previously obtained expansions of the bending angles\cite{Iyer,ChakrabortySen,Eiroa}, 
and are confirmed to be consistent with them.
Both weak and strong deflection expansions for the bending angles now becomes available up to an arbitrary order.
The method used here can give another analytical tool to study similar problems in other geometries such as Reissner-Nordstr{\"{o}}m spacetime for 
an arbitrary value of charge or Kerr-Newman spacetime. 

By comparing the coefficients of the power series for the deflection angles Eqs.(\ref{Schwarzschildsolution}) and (\ref{eRNsolution}), 
it can be seen that the bending angle for the Schwarzschild spacetime is larger than that for the extremal Reissner-Nordstr{\"{o}}m spacetime,
for every fixed value of $M/b$.
This is due to the repulsive effect caused by electric charge as shown numerically in ref.\cite{Hsiao}.

It is known that there are several transformation formulas for Appell functions\cite{KoikeShiga,Matsumotoohara,Matsumoto,Otsubo} but we could not get an adequate transformation to relate the two expressions (\ref{Appelleq}) and (\ref{Schwarzschildsolution}).
It is very interesting if we can find an appropriate transformation to obtain the deflection angle not only for Schwarzschild geometry but also for Kerr-Newman geometry.

\section*{Acknowledgement}
The authors thank Kaito Nasu for helpful discussions.

%%%%%%%%%%%%%%%%%%%%%%%%%%%%%%%%%%%%%%%%%%%%%%%%%%%%%%%%%%%%%%%%%%%%%%%%
\end{document}